\documentclass[twocolumn]{revtex4-1}
\newcommand{\beq}{\begin{equation}}
\newcommand{\eeq}{\end{equation}}
\usepackage{graphics}
\usepackage[hmargin=2cm,vmargin=3cm]{geometry}
\usepackage{dcolumn}
\usepackage{bm}
\usepackage{amsmath}
\usepackage{amsmath}
\usepackage{amsfonts}
\usepackage{amssymb}
\usepackage{graphicx}
\usepackage{color}
\usepackage{bm}
\usepackage{dashrule}
\usepackage{ulem}

\begin{document}

%\title{A spin liquid state in dimeric Ir units of hexagonal Ba$_3$ZnIr$_2$O$_9$ triggered by competing spin orbit and crystal field effects and strong quantum fluctuations}
\title{Spin-orbital liquid state assisted by singlet-triplet excitation in $J$~=~0 ground state of Ba$_3$ZnIr$_2$O$_9$}
%\title{Spin-orbit coupling induced excitonic spin liquid state in the hexagonal $d$$^{4}$ iridate Ba$_3$ZnIr$_2$O$_9$ }

\author{Abhishek Nag,$^{1}$ %
Srimanta Middey,$^{2}$ %
Sayantika Bhowal,$^{3}$ %
Swarup Panda,$^{2}$ %
Roland Mathieu,$^{4}$ %
J. C. Orain,$^{5}$ %
F. Bert,$^{5}$ %
P. Mendels,$^{5}$ %
P. Freeman,$^{6,7}$ %
M. Mansson,$^{6}$ %
H. M. Ronnow,$^{6}$ %
M. Telling,$^{8}$ %
P. K. Biswas,$^{9}$ %
D. Sheptyakov,$^{10}$ %
S. D. Kaushik,$^{11}$
Vasudeva Siruguri,$^{11}$ %
Carlo Meneghini,$^{12}$ %
D. D. Sarma,$^{13}$ %
Indra Dasgupta,$^{2,3,{\star}}$ %
Sugata Ray$^{1,2,{\star}}$\\ Email: \href{}{sspid@iacs.res.in (theoretical), mssr@iacs.res.in (experimental)}}
%sspidg@iacs.res.in (theoretical), mssr@iacs.res.in (experimental)

\vspace{0.1cm}

\affiliation{$^1$Department of Materials Science,~Indian Association for the Cultivation of Science, Jadavpur, Kolkata~700032, India\\ %
$^2$Centre for Advanced Materials, Indian Association for the Cultivation of Science, Jadavpur, Kolkata~700032, India\\%
$^3$ Department of Solid State Physics, Indian Association for the Cultivation of Science, Jadavpur, Kolkata~700032, India\\ %
$^4$Department of Engineering Sciences, Uppsala University, P.O. Box 534, SE-751 21 Uppsala, Sweden\\ %
$^5$Laboratoire de Physique des Solides, UMR CNRS 8502, Université Paris-Sud, 91405 Orsay, France\\ %
$^6$Laboratory for Quantum Magnetism (LQM), \'{E}cole Polytechnique F\'{e}d\'{e}rale de Lausanne (EPFL), Station 3, CH-1015 Lausanne, Switzerland\\ %
$^7$Jeremiah Horrocks Institute for Mathematics, Physics \& Astrophysics, University of Central Lancashire, Preston PR1 2HE, United Kingdom\\ %
$^8$ISIS Facility, Rutherford Appleton Laboratory, Chilton, Didcot, Oxon OX110QX, United Kingdom\\ %
$^9$Laboratory for Muon Spin Spectroscopy, Paul Scherrer Institute, CH-5232 Villigen PSI, Switzerland\\ %
$^{10}$Laboratory for Neutron Scattering and Imaging, Paul Scherrer Institute, CH-5232 Villigen PSI, Switzerland\\ %
$^{11}$UGC-DAE-Consortium for Scientific Research Mumbai Centre, R5 Shed, Bhabha Atomic Research Centre, Mumbai 400085, India\\ %
$^{12}$Dipartimento di Scienze,~Universit\'{a} Roma Tre,~Via della Vasca Navale,~84 I-00146 Roma,~Italy\\ %
$^{13}$Solid State and Structural Chemistry Unit, Indian Institute of Science, Bangalore 560012, India}

\begin{abstract}
Strong spin-orbit coupling (SOC) effects of heavy $d$-orbital elements have long been neglected in describing the ground states of their compounds thereby overlooking a variety of fascinating and yet unexplored magnetic and electronic states, until recently. The spin-orbit entangled electrons in such compounds can get stabilized into unusual spin-orbit multiplet $J$-states which warrants severe investigations. Here we show using detailed magnetic and thermodynamic studies and theoretical calculations the ground state of  Ba$_3$ZnIr$_2$O$_9$, a 6$H$ hexagonal perovskite is a close realisation of the elusive $J$~=~0 state. However, we find that local  Ir moments are spontaneously generated due to the comparable energy scales of the singlet-triplet splitting driven by SOC and the superexchange interaction mediated by strong intra-dimer hopping. While the Ir ions within the structural Ir2O9 dimer prefers to form a spin-orbit singlet state (SOS) with no resultant moment, substantial interdimer exchange interactions from a frustrated lattice ensure quantum fluctuations till the lowest measured temperatures and stabilize a spin-orbital liquid phase.
\end{abstract}

\maketitle
%PACS number(s): 75.10.-b, 75.47.Lx

The electronic and magnetic properties of 5$d$ transition metal (TM) compounds often exhibit unusual properties due to the presence of strong spin-orbit coupling (SOC), which is comparable to their on-site Coulomb ($U$) and crystal field ($\Delta_{\rm CFE}$) interactions~\cite{soc_phasedia}. While quantum numbers $m_{\rm l}$ (orbital) and $m_{\rm s}$ (spin), for each electron can still be considered good and the spin-orbit coupled many-body multiplets can be denoted by spectroscopic term symbols in the weak spin-orbit ($L$-$S$) coupling limit, the representation gets strongly modified for heavy multielectron elements having strong SOC. In such $j$-$j$~coupling regime, total $M_J$ ($\Sigma m_j$) becomes the only valid quantum number and the multiplets and their degeneracies are all solely determined by the total angular momentum $J$.

The electronic and magnetic response of a system in the $j$-$j$ coupling limit are not yet well-understood and therefore, in the recent time, has generated significant curiosity. For example, the unconventional insulating state of the layered tetravalent iridates (Ir$^{4+}$; 5$d^5$), designated as novel $j_{eff}$ = $\frac{1}{2}$ Mott insulators, have attracted much interest~\cite{kim_prl,noh_prl,arima_science}, as metallic behavior is usually expected for partially filled broad $t_{2g}$ bands in iridates. The insulating behavior is explained within single particle theories by assuming splitting of $t_{2g}$ bands into a fully filled $J_{eff}$~=~3/2 quartet bands and half filled narrow $J_{eff}$~=~1/2 doublet bands due to finite SOC, which further splits into a fully occupied lower Hubbard and empty upper Hubbard band in the presence of relatively small Hubbard $U$.

Another intriguing case would be a pentavalent Iridium compound (Ir$^{5+}$; 5$d^4$), where all the spin-orbit entangled electrons will be confined to the singlet $J$~=~0 ($M_J$~=~0) state, without any resultant moment or magnetic response. This is shown in Fig. 1(a) where the energy levels  of a low spin 5$d$ $t_{2g}^4$ Ir$^{5+}$ ion calculated as a function of spin-orbit coupling parameter $\lambda'$ ~\cite{Matsuura} evolves from the $L$-$S$ coupling to the $j$-$j$ coupling regime~\cite{jce} and a $J$~=~0 ground state is realized under strong SOC (see Supplementary Information (SI) for details). Recently there has been a theoretical suggestion of an excitonic magnetism (transition from the $J$~=~0 to $J$~=~1 state) in such systems~\cite{khaliullin_prl}, due to comparable energy scales of the singlet-triplet splitting determined by SOC and the super-exchange interaction promoted by hopping. Therefore, if $d^{4}$ Ir compounds indeed fall into the category of the elusive $j$-$j$ coupled systems, experimentally a zero or vanishingly small magnetic moment should be observed for the otherwise nominal $S$~=~1 Ir$^{5+}$ center and it would be interesting to know the nature of their magnetic response(s), if any. However, stabilization of such an unusual electronic state may be subtle as this $J$~=~0 ground state at moderate $\lambda'$ could be exceptionally fragile to minute external perturbation and enhanced magnetic responses may appear with rather small variations in magnetic fields or IrO$_6$ octahedral distortions giving rise to noncubic crystal field~\cite{khaliullin_prl,Cao_prl}.

Possibly for this reason, there has been no experimental realization of this unusual state till date, other than a other than a speculation in the case of NaIrO$_3$~\cite{cava_jssc}, while another possible candidate Sr$_2$YIrO$_6$ double perovskite acquired long range magnetic order with comparatively larger paramagnetic moment of 0.91~$\mu_B$/Ir (ideal moment = 2.83~$\mu_B$) because of noncubic IrO$_6$ crystal field~\cite{Cao_prl}.

\begin{figure}
\centering
\resizebox{8.5cm}{!}
{\includegraphics[scale=0.5]{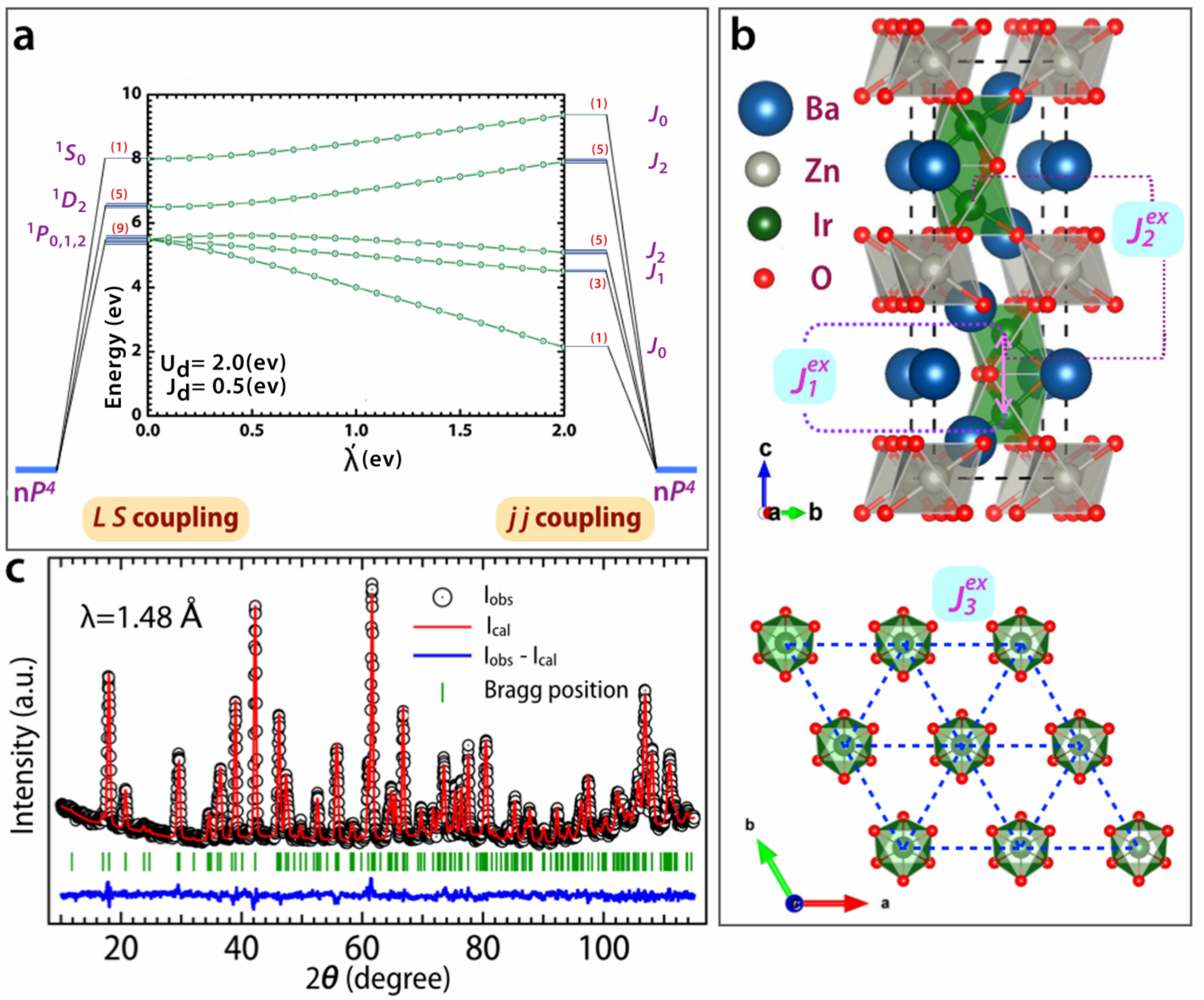}}
\caption{(Color Online)(a) The energy states of low spin 5$d$ $t_{2g}^4$ Ir$^{5+}$ atom calculated as a function of spin-orbit coupling parameter. (b) The  crystal structure of Ba$_{3}$ZnIr$_{2}$O$_{9}$ (upper panel) and triangular lattice formed by Ir ions  in $ab$ plane (lower panel). (c)  Experimental (black) and refined (red) neutron powder diffraction patterns at room temperature.}
\end{figure}

In contrast to  previous attempts, we have explored the complex electronic and magnetic interplay of Ir$^{5+}$ within a 6$H$ hexagonal compound, namely Ba$_3$ZnIr$_2$O$_9$ (Space group:P{\rm 6}$_{3}$/mmc; See Fig. 1(b)), where instead of the octahedral corner sharing of the IrO$_6$ units, the octahedra share face and form Ir$_2$O$_9$ dimers. Our experimental study indeed reveal that Ba$_3$ZnIr$_2$O$_9$ probably comes closest to the coveted $J$~=~0 ground state with a low magnetic moment of individual Ir atoms in the paramagnetic regime, indicating presence of rather strong $\lambda'$, but not strong enough to completely arrest the hopping assisted excitonic magnetism at low temperatures. Interestingly, first principles density functional calculations indicate that the adjacent Ir octahedra in a dimer prefers to be in a spin-orbital singlet (SOS) state which should have eliminated any net magnetization from each Ir$_2$O$_9$ dimeric unit provided this intradimer singlet interaction is strong enough. Instead, sufficient interdimer hoppings are present in order to induce quantum fluctuations in the SOS state and due to frustration arising from the triangular lattice structure in the $ab$-plane (lower panel of Fig. 1(b))~\cite{khaliullin_prl} possibility of any long range order is prohibited. Finally, detailed muon spin-rotation ($\mu$SR) and heat capacity measurements reveal the presence of strong quantum fluctuations at low temperatures, and absence of any long range magnetic order down to 100~mK suggesting  Ba$_3$ZnIr$_2$O$_9$ to be a rare example of a quantum  spin-orbital liquid (QSOL).

Polycrystalline Ba$_3$ZnIr$_2$O$_9$ (BZIO) was synthesized by standard solid state reaction using stoichiometric amounts of BaCO$_3$, ZnO and Ir-metal as starting materials~\cite{FeRu_own}.  The sample purity was checked and refined by powder x-ray diffraction measured in the Indian beamline (BL-18B) at Photon Factory, KEK, Japan. Neutron powder diffraction (NPD) patterns, recorded in National Facility for Neutron Beam Research (NFNBR), Dhruva reactor, Mumbai (India) and Paul Scherrer Institute (PSI), Switzerland, both of comparable resolution, were refined by Rietveld method using FULLPROF~\cite{fullprof}.  The x-ray photoelectron spectroscopic (XPS) measurements were carried out in an Omicron electron spectrometer, equipped with EA125 analyzer. $\mu$SR experiments were mostly performed with the EMU spectrometer at the ISIS large scale facility both in a helium flow cryostat and a dilution fridge. A few additional runs were measured with the DOLLY spectrometer at the PSI continuous source facility where the fast early time relaxation at low temperatures can be best resolved. The $dc$ magnetic measurements and heat capacity measurements were carried out using a Quantum Design SQUID magnetometer and a Quantum Design PPMS (physical property measurement system), respectively. X-ray absorption spectra at the Zn-$K$ and Ir $L_3$-edges were collected at the Elettra (Trieste, Italy) 11.1R-EXAFS beamline in standard transmission geometry at room temperature. All the electronic structure calculations were carried out using density functional theory (DFT) within generalised gradient approximation (GGA) using Hubbard $U$ and SOC as implemented in Vienna ab-initio simulation package (VASP) ~\cite{vasp1,vasp2}. This method uses a plane-wave basis set along with projector augmented waves (PAW) in the ionic core region ~\cite{PAW1,PAW2}.

Powder XRD data collected from polycrystalline BZIO (not shown) confirm complete phase purity. The NPD pattern (with $\lambda$ = 1.48 \AA; Fig. 1(c)) recorded at room temperature has been refined with space group $P6_3/mmc$ which revealed $<5\%$ Zn/Ir site-disorder, {\it i.e.} face shared octahedral positions (4$f$ sites) are almost exclusively occupied by Ir forming Ir$_2$O$_9$ dimer, while Zn ions solely occupy the isolated octahedral sites (2$a$). The structural parameters obtained from the refinement of the 1.5K data have been listed in Supplementary Table I of SI. However, the local cationic disorder and octahedral distortions in these compounds may pose a different picture~\cite{FeRu_own}, and therefore XAFS experiments were carried out. The  parameters, obtained from Zn $K$- and Ir $L_3$-XAFS data analysis are summarized in Table I. The XAFS data analysis confirms negligible chemical disorder with Ir (Zn) ions mainly located at the $4f$ ($2a$) sites with local interatomic distances being consistent with literature. The average Ir-O distance is compatible with the average crystallographic Ir-O distance (1.98 \AA). It must be noted that the $\sigma^2_{IrO}$ is comparable with (and even lower than) the Zn-O one, which suggests that ZnO$_6$ and IrO$_6$ octahedra have similar distortions.

Next, the temperature dependence of the electrical resistivity ($\rho$) of BZIO, measured by standard four probe method, is shown in the inset to Fig. 2(a). The insulating behavior can be modeled by variable range hopping mechanism in 2-dimension as shown in Fig. 2(a). The insulating behavior has been further tested by measurement of valence band photoemission experiment, where absence of any density of states at the Fermi level (see Fig.2(b)) is confirmed. The XPS spectra for the Ir 4$f$ core level (Fig. 2(c)) can be fitted by spin-orbit split doublet with separation of 3.04 eV. The energy positions of 4$f_{5/2}$ and 4$f_{3/2}$ doublet confirm +5 oxidation state of Ir~\cite{Ir_4f}.

This intriguing insulating behavior of Ir$^{5+}$ having four $d$-electrons in low spin configuration has been investigated within single particle mean-field framework by {\it ab-initio} density functional theory calculations by systematically including Hubbard $U$ and SOC. The non-spin polarized density of states (DOS) of BZIO within GGA is shown in Fig. 2(d). As expected, the system is metallic with the Fermi level located at the $t_{2g}$ manifold, comprising of twelve bands from four Ir atoms in the unit cell. The octahedral crystal field is strong, resulting in quite large ($\sim$3.5 eV) $t_{2g}$ - $e_{g}$ crystal field  splitting. The strong intradimer hoppings further lifts the degeneracy of the $t_{2g}$ states. The DOS is however remarkably different upon inclusion of SOC and a Hubbard $U$ ($U$-$J$= 2.5 eV) which makes the system insulating with a gap ~10 meV (see inset of Fig 2(e)). In the presence of SOC the $t_{2g}$ states (six states including spin degeneracy) break into a spin-orbit entangled set of states namely two-fold degenerate $\Gamma_{7} $ and a four-fold degenerate $\Gamma_{8}$ states. Since four electrons are available per Ir, all the $\Gamma_{8}$ states are occupied and the system is insulating as seen experimentally.

\begin{figure}
\centering
\resizebox{8.5cm}{!}
{\includegraphics[scale=0.5]{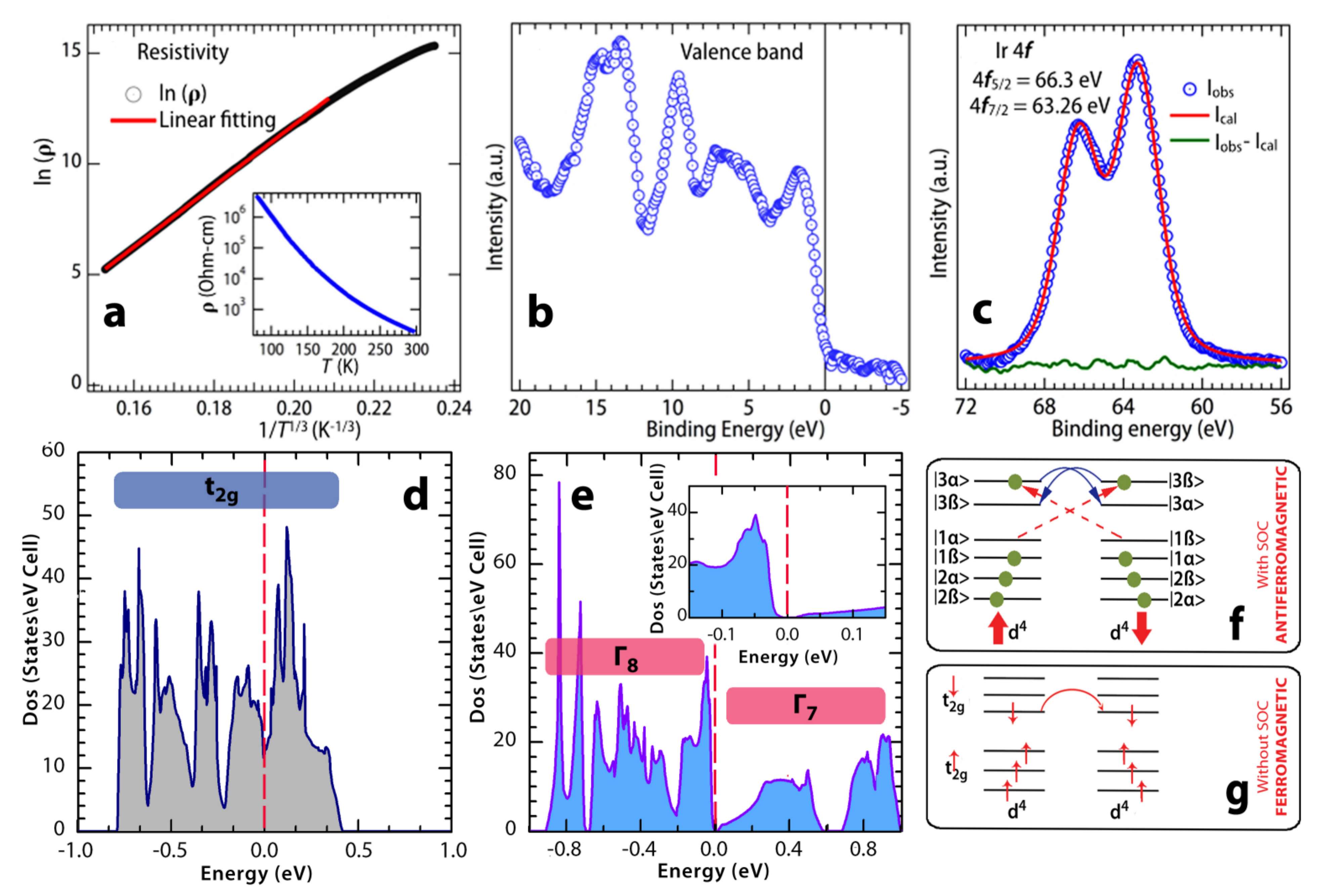}}
%%{\includegraphics{Fig2}}
\caption{(Color Online) (a) The temperature dependence of resistivity (inset) can be described by variable range hopping in two dimension (main panel). XPS spectra for (b) valence band and (c)Ir 4$f$ core level, recorded with Mg $K_\alpha$ radiation. Total density of states obtained within (d) GGA and (e) GGA+SOC+U. Mechanism of (f) antiferromagnetic interaction in presence of SOC and (g) ferromagnetic interaction in absence of SOC.}
\end{figure}
While analyzing the magnetic properties of Ba$_{3}$ZnIr$_{2}$O$_{9}$ in the framework of GGA+U+SOC we find that local moments are spontaneously generated in the magnetic phase due to the hybridization between the occupied $\Gamma_{8}$ and empty  $\Gamma_{7} $ states. In the following we shall argue that the superexchange between a pair of Ir atoms within a dimer will promote a spin-orbital singlet state.  The four $\Gamma_{8}$ orbitals are designated as $\vert 1 \alpha \rangle$, $\vert 1 \beta \rangle$, $\vert 2 \alpha \rangle$, $\vert 2 \beta \rangle$ while the two $\Gamma_{7}$ orbitals as $\vert 3 \alpha \rangle$, $\vert 3 \beta \rangle$ as illustrated in Fig. 2(f). We regard these states as pseudo-spin states, {\it i.e.} the state in which one electron occupying $\vert i \alpha \rangle$ ($\vert i \beta \rangle$) will be considered as up (down) pseudo spin states. It is interesting to note that unlike $t_{2g}$ orbitals the hopping between the pseudo spin states are not always pseudo spin conserved~\cite{Matsuura2}. In particular, an electron in a pseudo spin state $\vert 1 \alpha \rangle$ ($\vert 1 \beta \rangle$) can hop only to pseudo spin states $\vert j \beta \rangle$ ($\vert j \alpha \rangle$) where $j \neq 1$ and all other hoppings are pseudo spin conserved~\cite{Matsuura2}.

Such hopping leads to antiferromagnetic pseudo spin interaction and thereby favors SOS state as shown schematically in Fig. 2(f). For an antiferromagnetic configuration of pseudo spins in a dimer the $\Gamma_{8}$ electron in the state $\vert 1 \alpha \rangle$ and  $\vert 1 \beta \rangle$ can virtually hop to  $\vert 3 \beta \rangle$ and $\vert 3 \alpha \rangle$ respectively as this hopping is not pseudo-spin conserved (indicated by $\dashrightarrow$ in Fig. 2(f). Once these electrons are promoted to the $\Gamma_{7}$ state, they can then gain energy by hopping to the other $\Gamma_{7}$ level as this hopping is now pseudo-spin conserved (indicated by ---- in Fig. 2(f)). The latter hopping is however not possible if the pseudo spins residing in the dimer were parallel. The same dimer however favors ferromagnetic interaction in the absence of spin-orbit interaction as illustrated in Fig. 2(g) as the hopping between the $t_{2g}$ orbitals are spin conserved. In order to check the above mentioned scenarios we have carried out spin polarized GGA+U and GGA+U+SOC calculations for several magnetic configurations (See SI). The GGA+$U$+SOC calculations reveal a rather large spin moment (0.65-0.75 $\mu$B) per Ir atom. In addition, the large value of the orbital moment (0.26 $\mu$B) confirms the importance of spin-orbit coupling. Also, the magnetic configuration with antiferromagnetic intradimer coupling is found to be energetically favorable only in the presence of SOC, emphasizing its importance in the realization of SOS states in BZIO.  Although, an estimate of symmetric exchange interactions suggest appreciable intradimer exchange ($J_{1}$ = -14.6 meV), there is substantial interdimer exchange too ($J_{3}$ = -1.5 meV with 6 neighbors in the $a-b$ plane) and ($J_{2}$ = -1.6 meV with 3 neighbors perpendicular to the $a-b$ plane). Therefore, the intradimer interactions are not strong enough compared to the interdimer interactions leading to a frustrated behavior instead of a SOS-like long-range state.

In order to check whether the SOC driven SOSs  in BZIO eventually lead to long range magnetic order, we have carried out NPD experiments at lower temperature too. The high resolution NPD pattern recorded at 2~K has been shown in Fig. 3(a). All peaks of the observed pattern have been refined with  $P6_3/mmc$ space group without any significant magnetic contribution, indicating absence of any long range magnetic order in the system at least down to 2~K. These NPD results also rule out the possibility of any structural transition below 300~K. However, in presence of strong SOC and frustration, a clear long range magnetic order might be missingabsent while local magnetic interactions are still significant. Therefore, in order to unravel the nature of the ground state stabilized in this compound we performed muon spin relaxation experiments in zero field, a technique perfectly suited to detect weak and/or partial freezing of local magnetic moments. The evolution of the muon polarization in the sample is shown in Fig. 3(b). At high temperature, the slow Gaussian like relaxation arises from nuclear dipoles which are static in the $\mu$SR time window and give a $T$ independent contribution. Upon lowering temperature, the relaxation rate increases gradually from about 100~K as a result of the slowing down of the electronic fluctuations and strikingly levels off below about 2~K as demonstrated by the nearly perfect overlap of the 2~K and 0.1~K polarization curves. Despite this obvious slowing down of the dynamics, we did not observe any spontaneous oscillations or emergence of a ``1/3rd" tail that would signal a magnetic transition to a frozen state. The robustness of the relaxation to the application of a longitudinal field at base temperature also supports its dynamical origin (see SI). Therefore, it can be concluded that the system does not order down to 0.1~K at least.

\begin{figure}
\centering
\resizebox{8.5cm}{!}
{\includegraphics[scale=0.5]{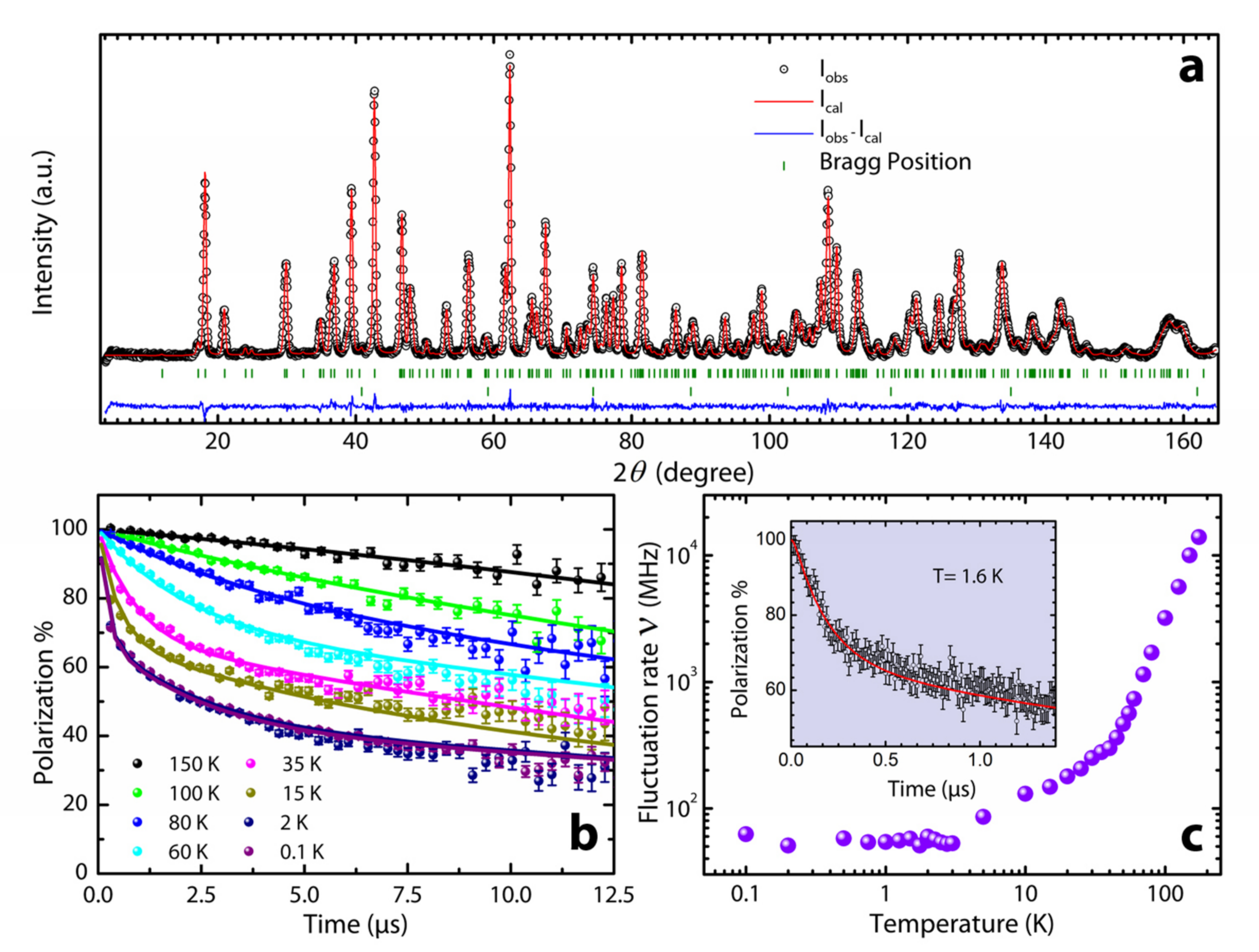}}
\caption{(Color Online) (a) High resolution neutron powder diffraction recorded at 2 K. (b) Time evolution of the muon polarization in Ba$_{3}$ZnIr$_{2}$O$_{9}$ in zero external field at different temperatures. Lines are fits of the data described in the text. (c) Fluctuation rate of the internal fields versus temperature as extracted from the fits of the polarizations (see text). Inset: At low temperature the fast initial relaxation of the muon polarization as measured on the DOLLY spectrometer at the continuous muon source of the Paul Scherrer Institut.}
\end{figure}

In the low temperature regime, below about 35K, the signal appears clearly to be composed of at least 3 components; one fast relaxing, one with a moderate relaxation rate and one hardly relaxing. These three components likely reflect three different muon stopping sites in the sample. Therefore we fitted the data to the model,

\begin{equation}
\begin{split}
P(t)~=~f_{fast}G(t, \triangle H_1, \nu)~+~f_{slow}G(t, \triangle H_2, \nu)\\
+~(1~-~f_{fast}~-~f_{slow})KT(t, \triangle H)
\end{split}
\end{equation}

where $KT$ is the Kubo-Toyabe function~\cite{kt} accounting for the static nuclear field distribution of width $\triangle H$, $f_{fast}$ and $f_{slow}$ are the fractions of the two more relaxing components and $G(t, \triangle H, \nu)$ is the dynamical relaxation function introduced in Ref. 19

to account for the effect of a fluctuating field distribution of width $\triangle H$ at a rate $\nu$. The fits with respect to this model are shown in Fig. 3(b) where only the fluctuation rate $\nu$ was allowed to vary with temperature and the extracted values are given in Fig. 3(c). The other $T$ independent parameters were refined to $\triangle H~=~0.4(2) G, \triangle H_1~=~147(5) G, \triangle H_2~=~39(3) G$, $f_{fast}~=~f_{slow}~=~31(2)\%$. At the lowest measured temperatures, when the fluctuations are the slowest we note that for both relaxing components the inequality $\nu > \gamma\triangle H$ where $\gamma$~=~2$\pi$~x~135.5 Mrad/s/T is the muon gyromagnetic ratio, still holds since $\gamma\triangle H_1$~=~12.5$\mu$s$^{-1}$ and $\gamma\triangle H_2$~=~3.4$\mu$s$^{-1}$. This validates the use of dynamical relaxation functions, in line with the absence of the usual evidence for a frozen ground state.

The above prediction of a special QSOL state in BZIO is further confirmed by the analysis of heat capacity ($C$). The temperature variation of C (Fig. 4(a)) does not show any peak or anomaly, consistent with the absence of any structural and/or long range magnetic transition. In order to extract the magnetic heat capacity $C_m$ by subtracting the lattice contribution, $C$ was also measured for an isostructural nonmagnetic compound Ba$_3$ZnSb$_2$O$_9$.  The difference of molecular weight between these two compounds for subtracting the lattice part has been taken into account following the scaling procedure, developed by Bouvier {\it et. al.}~\cite{Bouvier}. $C_m$ obtained from such analysis has been plotted in Fig. 4(b), which shows only a broad peak around $\sim$15 K followed by a decay around 30~K without any further anomaly. A fit of the magnetic heat capacity ($C_m$=$\gamma T + \beta T^2 $) from 3K to 9K gives a significant $T$-linear contribution of $\gamma$~=~25.9 mJ/mol-$K^2$ unusual for insulating systems. Also, the finite $\gamma$ value does not change with application of external magnetic field as high as 9 Tesla, suggesting its origin to be gapless excitations from spinon density of QSOL state or minor lattice oxygen vacancies and not any paramagnetic impurity~\cite{gaplessHC,spinon_paramagnet,spinon,spinon3,gamma_vacancy}. This mixed temperature dependence is observed in other spin liquid systems, {\it e.g.} Ba$_3$CuSb$_2$O$_9$ ($S$~=~1/2)~\cite{cusb2}, 3C phase of Ba$_3$NiSb$_2$O$_9$ ($S$~=~1)~\cite{nisb2}, NiGa$_2$S$_4$ ($S$~=~1)~\cite{niga2}, Ba$_3$IrTi$_2$O$_9$ ($J_{eff}~=~1/2$)~\cite{mahajan2012} {\it etc.} The magnetic entropy obtained by integrating $C_m$/$T$ with $T$ has been shown in Fig. 4(c), which is much smaller ($\sim$7\%) compared to the value for spin only magnetic entropy, Rln(2$S$+1) with $S$~=~1 and unlike other spin liquid candidates containing 3$d$ TM, where the entropy change is more than 30-50\% of the total magnetic entropy~\cite{cusb2,nisb2,ramirez}. The value is in fact close to 0 as expected for a spin-orbit coupled magnetic entropy, Rln(2$J$+1) with $J$~=~0.  Therefore, the heat capacity study does confirm presence of QSOL-like dynamical ground state in this system, visible below 30~K. It also indicates towards the fact that the magnetic ground state and the moment in BZIO are only `byproducts' of certain spin excitations from otherwise nonmagnetic, singlet Ir$^{5+}$ $J$~=~0 state, since the small magnetic entropy is likely released due to such weak magnetic crossover~\cite{Cao_prl}.

Finally, we show the temperature dependence of field-cooled (FC) and zero-field-cooled (ZFC) magnetization measured with 5000~Oe field in Fig.~4(d). Clearly there is no indication of a magnetic order in this case, as expected. These $\chi$ vs. $T$ curves are fit with a function, $\chi_0$ + $C_W$/($T$ - $\theta_W$) where $\chi_0$ is the temperature independent paramagnetic susceptibility, $C_W$ and $\theta_W$ represent Curie constant and Curie-Weiss temperature, respectively, and the result is shown as a green straight line, laid on the 1/($\chi$-$\chi_0$) vs. $T$ data. The fitting, especially at the higher temperature region, extracts a $\theta_W$ value $\sim$-30 K and a moment of $\sim$0.2~$\mu_B$/Ir. Clearly, the magnetic frustration parameter, defined as the ratio of $\theta_W$~=~-30~K to the upper limit of the magnetic ordering temperature that we could probe, going upto at least 300 proves that the system indeed behaves like a quantum spin-orbital liquid.

\begin{figure}
\centering
\resizebox{7cm}{!}
{\includegraphics[scale=0.5]{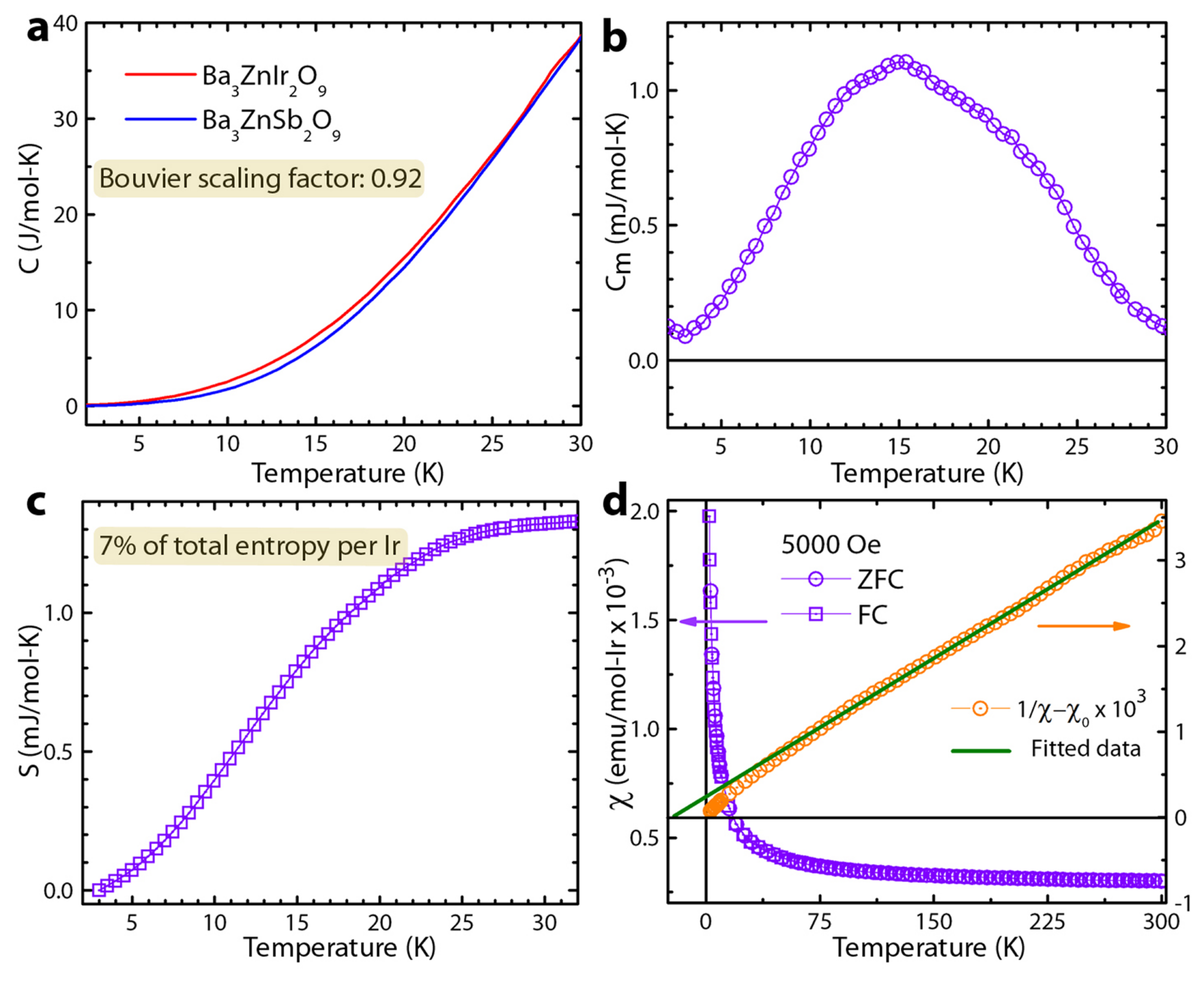}}
\caption{ (Color Online) Temperature dependence of (a) specific heat ($C$), (b) magnetic specific heat ($C_m$) after subtracting the lattice contribution. (c) Magnetic entropy as a function of temperature. (d) FC-ZFC magnetization measured at 5000 Oe magnetic field fitted using Curie-Weiss law. 1/($\chi$-$\chi_0$) vs. $T$ is also plotted showing $\theta_W$.}
\end{figure}

In conclusion, our detailed analysis reveal BZIO is a $J=0$ Insulator, where the origin of the small but finite  nonzero moment is due to super-exchange induced spin excitations~\cite{khaliullin_prl}.
These Ir moments within the Ir$_2$O$_9$ dimer interact antiferromagnetically resulting in spin-orbital singlets. The inter-dimer hoppings in the $ab$-plane induce quantum fluctuations in these singlets and a long range magnetic order is arrested due to  geometric frustration, arising from the triangular lattice structure in the $ab$-plane resulting in a quantum spin-orbital liquid state, confirmed both by $\mu$SR and heat capacity studies.

\section{Acknowledgements}
A.N. and S.M. thanks CSIR, India for a Fellowship. SR thanks CSIR, India for funding (project no. (1269)/13/EMR-II) and Saha Institute of Nuclear Physics, India for facilitating the experiments at the Indian Beamline, Photon Factory, KEK, Japan. J.C.O., F.B. and P.M. thank the French Agence Nationale de la Recherche for its funding (grant "SPINLIQ" No. ANR-12-BS04-0021). Neutron diffraction work presented in this manuscript is partially based on experiments performed at the Swiss spallation neutron source SINQ, Paul Scherrer Institute, Villigen, Switzerland. The authors thank Indo-Swiss Joint Research Programme, Swiss National Science Foundation and its Sinergia network Mott Physics Beyond the Heisenberg Model (MPBH).The authors thank Dr. Avinash Mahajan for useful discussions.
\section{Author contributions}
A.N. and S.M. have contributed equally to this work.
\section{Additional information}
Supplementary information is available in the online version of the paper. Correspondence should be addressed to S.R. and I. D. G.

\begin{centering}
\begin{table*}[!h]
\caption {(Main text)Main structural parameters obtained from the Multi-shell fitting of the Zn $K$-edge and Ir $L_3$ edge XAFS spectra. The crystallographic distances, as obtained by ND Rietveld refinement, are reported for shake of comparison.}
%\begin{tabular}{|c|c|c|c|c|}
\begin{tabular}{lcccccclcccc}
\hline
\hline
            &  & Zn $K$ &  & & & & & &  Ir $L_3$  & &  \\  \cline{1-5}  \cline{8-12}
 Shell     & N        &  R    & $\sigma^{2}$    & R$_{\rm{NPD}}$ & & &Shell     & N        &  R    & $\sigma^{2}$    & R$_{\rm{NPD}}$  \\
             &            & ({\AA}) &                              ({\AA}$^{2}$)               & ({\AA}) & &  &          &            & ({\AA}) &                              ({\AA}$^{2}$)               & ({\AA})\\ \hline
Zn-O                       & 6                       & 2.05(1)                     & 6E-03                      & 2.08    &  & & Ir-O                        & 6                       & 1.96(1)                      & 5.4E-03                      & 1.9-2.0                      \\
              &            &                                                                     &              &              &  & & Ir-Ir                       & 1                       & 2.77(2)                      & 4.1E-03                      & 2.75                      \\
Zn-Ba                      & 8                       & 3.55(2)                     & 9.0E-03                      & 3.58 & &  & Ir-Ba                       & 7                       & 3.44(2)                      & 1.4E-02                      & 3.52                   \\
Zn-Ir                      & 6                       & 3.98(4)                     & 1.0E-02                      & 4.00    &  & & Ir-Zn                       & 3                       & 4.01(4)                      & 1.2E-02                      & 4.00                         \\
                        \\ \hline
\end{tabular}
\end{table*}
\end{centering}

\newpage

\pagebreak
\widetext
\begin{center}
\textbf{\large Supplementary information for `Spin-orbital liquid state assisted by singlet-triplet excitation in $J$~=~0 ground state of Ba$_3$ZnIr$_2$O$_9$'}
\end{center}
\vspace{.25 cm}

\setcounter{equation}{0}
\setcounter{figure}{0}
\setcounter{table}{0}
\setcounter{page}{1}

\maketitle
\begin{flushleft}
 \bf{1. Derivation of the energy level diagram for $t_{2g}^4$ configuration in $L-S$ and $j-j$ coupling scheme}
\end{flushleft}
\vspace{.25 cm}

To understand the energy levels of a single $t_{2g}^4$ state, let us consider the following three-orbital model Hamiltonian\cite{Matsuura}:
 \begin{equation}
  H_{tot}=H_d^{int}+H_d^{SO}
 \end{equation}
Where, $H_d^{int}$ and $H_d^{SO}$ are respectively the Hamiltonian for the Coulomb interaction and the spin-orbit interaction on d orbitals.
The Coulomb interaction can be written as,
\begin{equation}
   H_d^{int}= U_d \sum_{i=1,2,3} n_{i \uparrow}n_{i \downarrow}  \\
             +\frac{U_{d}'-J_d}{2} \sum_{\substack{i,i'=1,2,3 \\
                                          (i \neq i')}}
                                          \sum_{\sigma} n_{i\sigma}n_{i' \sigma} \newline
             + \frac{U_d'}{2} \sum_{\sigma \neq \sigma'}\sum_{\substack{i,i'=1,2,3 \\
                                                              (i \neq i')}}
                                                              n_{i\sigma}n_{i'\sigma '}
             + \frac{J_d}{2} \sum_{\substack{i,i'=1,2,3 \\
                                          (i \neq i')}}
                                          (d_{i' \uparrow}^\dagger d_{i \uparrow}d_{i \downarrow}^\dagger d_{i' \downarrow}+ h.c.)
\end{equation}
where, $U_d$, $U_{d}'$ and $J_d$ are respectively intra-Coulomb interaction, inter-Coulomb interaction and  Hund's rule coupling. These Coulomb interactions have the relation $U_d=U_d'+2J_d$. $d_{i\sigma}$($d_{i\sigma}^\dagger$) is the annihilation
(creation) operator of the $i$-th orbital $(i=1,2,3)$ with a spin $\sigma$ and $n_d=d_{i,\sigma}^\dagger d_{i,\sigma}$. \\

The explicit form of spin-orbit interaction is given by,

\begin{equation}
  H_d^{SO}= \frac{i\lambda'}{2} \sum_{lmn} \epsilon_{lmn} \sum_{\sigma \sigma'} \sigma_{\sigma \sigma'} ^n d_{l\sigma}^\dagger d_{m\sigma'}
\end{equation}
where, $\lambda'$ is the magnitude of spin-orbit interaction between orbital ($l_i$) and spin ($s_i$) angular momenta of the i-th electron
and $\epsilon_{lmn}$ is the Levi-Civita symbol.

For $d^4$ filling, the total number of possible configurations are $^6C_4=15$.
The eigenvalues and eigenstates are calculated by diagonalizing the Hamiltonian $H_{tot}$ in the above mentioned basis set. The variation of
the energy eigenvalues with $\lambda'$ for $U_d'=1.0$ eV and $J_d=0.5$ eV is shown in Fig 1(a) of the paper.
%\newpage
\maketitle
\begin{flushleft}
 \bf{2. Structural parameters after Rietveld refinement of NPD spectrum of Ba$_3$ZnIr$_2$O$_9$ at 1.5K.(Table. I)}
\end{flushleft}
\vspace{.25 cm}
\begin{centering}
\begin{table} [h]
\caption{(Supplementary Information) Structural data and agreement parameters\protect\footnotemark[1]}
\begin{tabular}{|c|c|c|c|c|c|c|}
\hline
Atom & Site & $x$  & $y$  & $z$  & $B$({\AA}$^2$) & Occ.\\
\hline
Ba1 & 2b & 0 & 0 & 1/4 & 0.23(3) & 1\\
\hline
Ba2 & 4f & 1/3 & 2/3 & 0.5876(1) & 0.26(2) & 1\\
\hline
Zn & 4f & 0 & 0 & 0 & 0.28(3) & 1\\
\hline
Ir & 2a & 1/3 & 2/3 & 0.15411(5) & 0.34(1) & 1\\
\hline
O1 & 6h & 0.4848(2) & 0.9696(5) & 1/4 & 0.44(1) & 1\\
\hline
O2 & 12k & 0.1701(2) & 0.3402(5) & 0.08318(6) & 0.60(1) & 1\\
\hline
\end{tabular}
\footnotetext[1]{Space group: $P6_3/mmc$, $a$~=~5.77034(9)~{\AA}, $c$~=~14.3441(2)~{\AA}, V~=~413.62(1)~{\AA}$^3$, $\chi^2$~=~1.38, $R_p$~=~7.04\%, $R_{wp}$~=~7.71\%, $R_{exp}$~=~6.57\%.}
\end{table}
\end{centering}

\begin{flushleft}
 \bf{3. Details of the Density Functional Theory Calculations}
\end{flushleft}
\vspace{.25 cm}
%\section{Details of the Density Functional Theory Calculations}
All  calculations reported in this work are carried out using a plane wave based method
as implemented in Viena \textit{ab-initio} simulation package (VASP)~\cite{vasp1,vasp2}. Exchange and correlation effects are treated
using local (spin) density approximation (LSDA) with generalized gradient correction (GGA) of Perdew-Burke-Ernzerhof
including Hubbard $U$~\cite{HubbardU} and spin-orbit coupling (SOC). A nominal value of $U-J$ = 2.5 eV has been taken unless otherwise stated
to analyze the effect of correlation on the electronic and magnetic properties of this system.
We have used projector augmented wave potentials~\cite{PAW1,PAW2} to model the electron-ion interaction.
The kinetic energy cut off of the plane wave basis was chosen to be 500 eV and 10$\times$10$\times$6 k-mesh has been
used for Brillouin-Zone integration. Symmetry has been switched off in order to minimize possible numerical
errors.

\begin{flushleft}
 \bf{4. Pseudo spin states}
\end{flushleft}
\vspace{.25 cm}
%\section{Pseudo spin states}
The four $\Gamma_8$ orbitals are given by,
 \begin{eqnarray*}
   |1\alpha\rangle &=& \frac{1}{\sqrt{2}}(|d_{yz\uparrow}\rangle+i|d_{zx\uparrow}\rangle),\\%$\linebreak$
   |1\beta\rangle &=& \frac{1}{\sqrt{2}}(|d_{yz\downarrow\rangle}-i|d_{zx\downarrow}\rangle),\\%$\linebreak$
   |2\alpha\rangle &=& \frac{1}{\sqrt{6}}(2|d_{xy\uparrow}\rangle-|d_{yz\downarrow\rangle}-i|d_{zx\downarrow}\rangle),\\%$\linebreak$
   |2\beta\rangle &=& \frac{1}{\sqrt{6}}(2|d_{xy\downarrow}\rangle+|d_{yz\uparrow}\rangle-i|d_{zx\uparrow}\rangle),
 \end{eqnarray*}
and the two $\Gamma_7$ orbitals are,
 \begin{eqnarray*}
  |3\alpha\rangle &=& \frac{1}{\sqrt{3}}(|d_{xy\uparrow}\rangle+|d_{yz\downarrow}\rangle+i|d_{zx\downarrow}\rangle),\\%$\linebreak$
  |3\beta\rangle &=& \frac{1}{\sqrt{3}}(|d_{xy\downarrow}\rangle-|d_{yz\uparrow}\rangle+i|d_{zx\uparrow}\rangle)
 \end{eqnarray*}

%\begin{abstract}
%Anything
%\end{abstract}

%\end{abstract}
\begin{flushleft}
 \bf{5. Results of spin polarized calculations}
\end{flushleft}
\vspace{.25 cm}

\begin{figure}[h]
\centering
\includegraphics[scale=0.2]{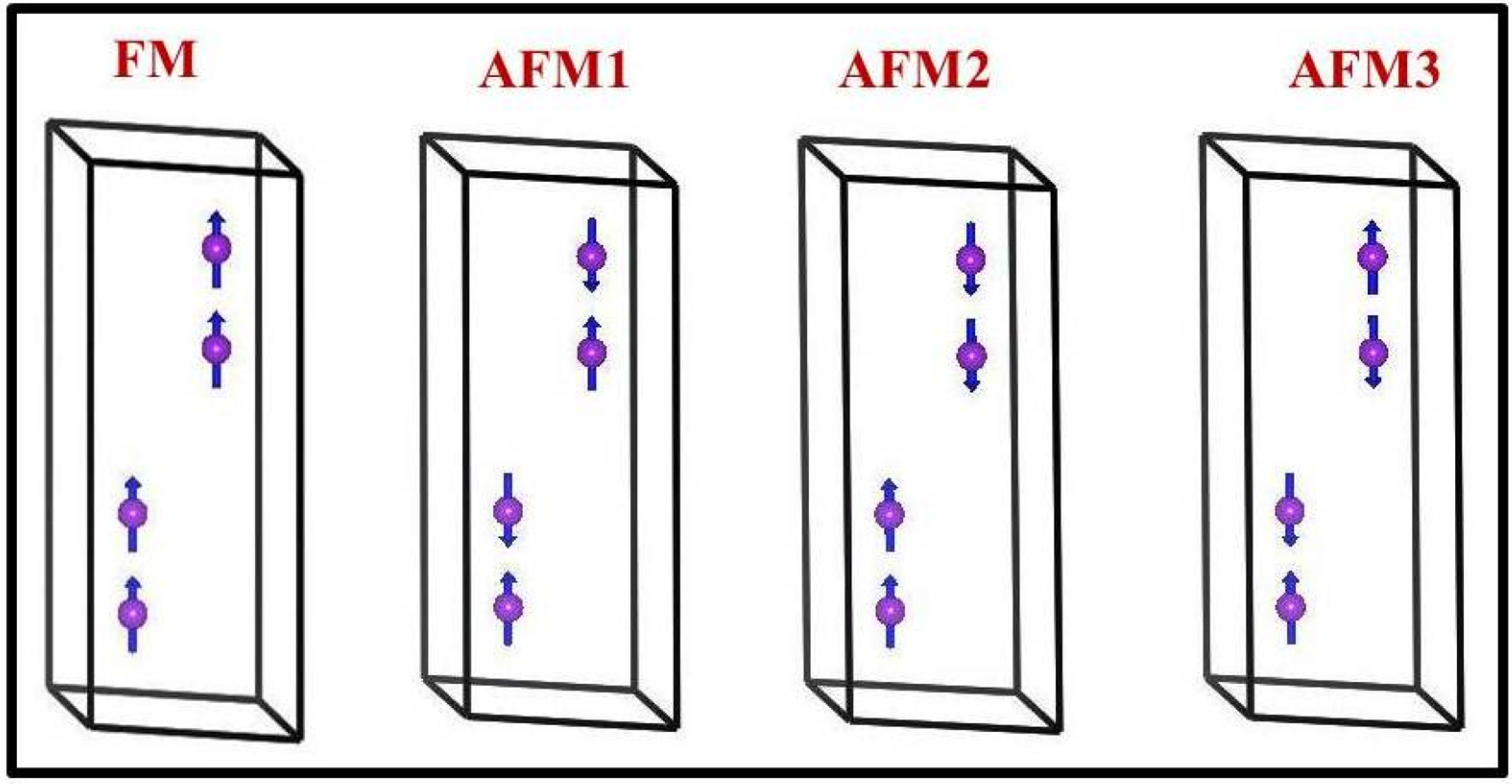}
\caption{The four possible magnetic states.}
\label{magnetic_state}
\end{figure}

%\section{Results of spin polarized calculations}
Four different magnetic configurations (see Fig.~\ref{magnetic_state})
namely FM (both the intra and inter dimer couplings are ferromagnetic), AFM1 (both the intra and inter dimer
couplings are antiferromagnetic), AFM2 (intra dimer coupling is ferromagnetic and inter dimer coupling is antiferromagnetic), and AFM3 (intra dimer
coupling is antiferromagnetic and inter dimer coupling is ferromagnetic) have been simulated to find out the lowest energy state.
Our calculations as summarized in Table~\ref{table:energy_ggau} reveal that AFM2 state has the lowest energy within GGA+$U$ approximation.
Table~\ref{table:energy_ggau} also display the energies of various magnetic states along with spin and orbital moments, obtained within GGA+$U$+SOC approach.
We find that the SOC changes the lowest energy state from AFM2 to AFM1 as discussed in the text.

\begin{table*}[h]
\begin{center}
\tabcolsep 2.8pt
\small
\begin{tabular}{| c| c| c|c|c|c|c|c|c|c|c|}
\hline
&\multicolumn{5}{|c|}{GGA+$U$} &\multicolumn{5}{|c|}{GGA+$U$+SOC} \\[1 ex]
\hline
& Energy &\multicolumn{4}{|c|}{Spin Moment} & Energy &\multicolumn{4}{|c|}{Spin (Orb) Moment}\\
& (meV/Ir) &\multicolumn{4}{|c|}{($\mu_B$)} & (meV/Ir) &\multicolumn{4}{|c|}{($\mu_B$)} \\[1 ex]
& & Ir & O1 & O2 & Tot & & Ir & O1 & O2 & Tot\\
\hline
FM & 0.0 & 1.09 & 0.10 & 0.19 & 8.00 & 0.0 & 0.62 (0.10) & 0.05 & 0.12 & 4.61 \\[-.5ex]
%\raisebox{1.5ex}{Ba$_3$ZnIr$_2$O$_9$}
AFM1 & 44.97 & 0.78 & 0.00 & 0.13 & 0.00 & -41.51 & 0.73 (0.26) & 0.00 & 0.11 & 0.00\\
AFM2 & -6.59 & 0.98 & 0.10 & 0.14 & 0.00 & 1.22 & 0.65 (0.20) & 0.05 & 0.10 & 0.00 \\
AFM3 & 26.96 & 0.86 & 0.00 & 0.18 & 0.00 & -34.68 & 0.74 (0.22) & 0.00 & 0.14 & 0.00\\
\hline
\end{tabular}
\end{center}
\caption{(Supplementary Information) Energy of various magnetic states along with the spin and orbital moments within GGA+$U$ and GGA+$U$+SOC approach. Energy for FM state is assumed to be zero. Orbitals moments of Ir are written in the parenthesis.}
\label{table:energy_ggau}
\end{table*}
%\vspace{-0.2 in}

%\vspace{-0.3 in}
\newpage
\begin{flushleft}
 \bf{6. Effect of applied magnetic field on muon spin-rotation ($\mu$SR) results}
\end{flushleft}
\vspace{.25 cm}

\begin{figure}
\centering
\resizebox{8.5cm}{!}
{\includegraphics[scale=0.4]{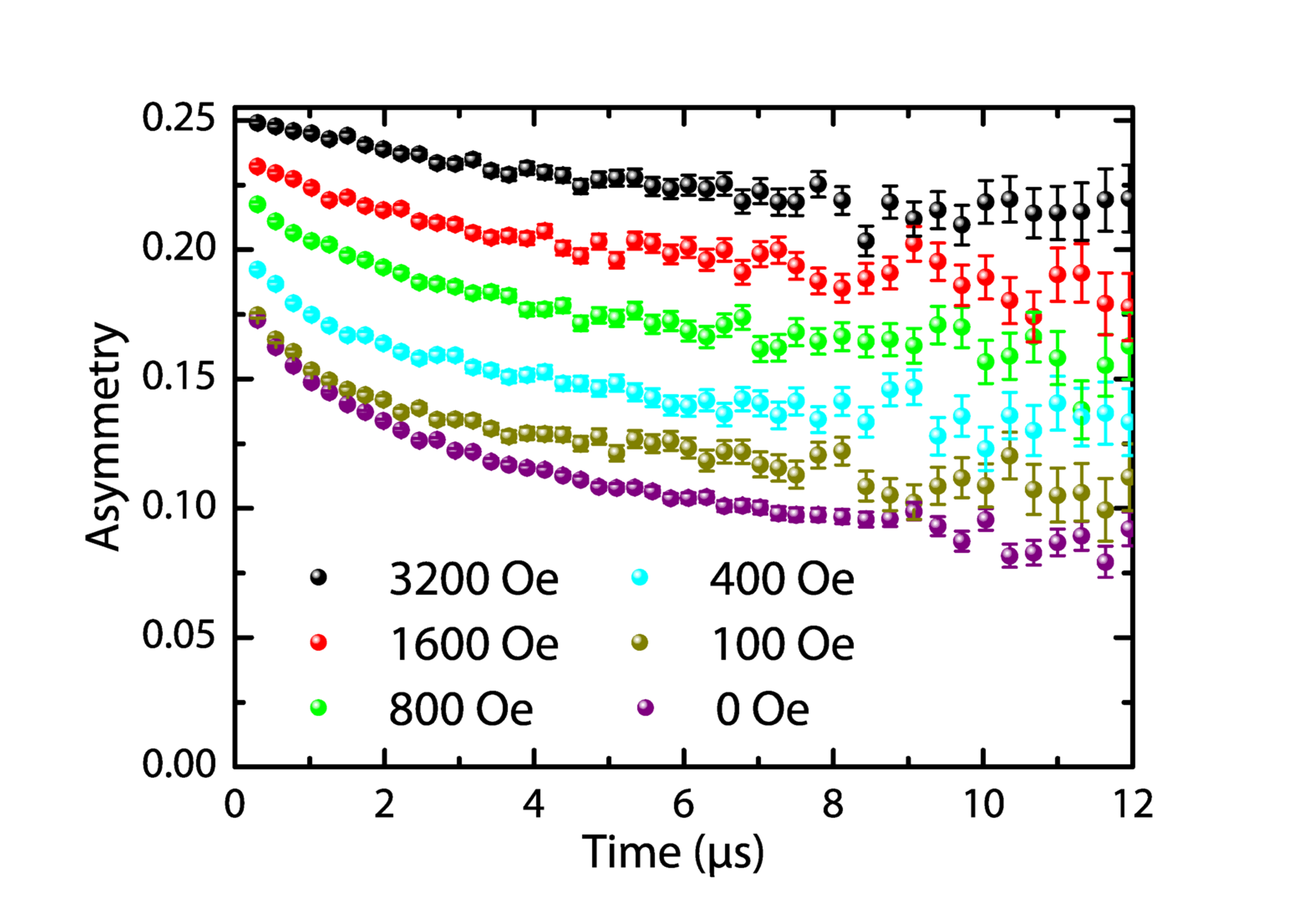}}
\caption{Asymmetries of the muon decay at 0.1 K measured with various longitudinal fields.}
\end{figure}

Fig. 2 shows the evolution of the relaxation when an external field is applied along the initial muon polarization at the lowest measured temperature 0.1~K. In case of a static field distribution of width $\triangle H_i$, a strong applied field $H_L$~5$\triangle H_i$ should overcome the effect of the internal field and almost completely suppress the relaxation of the muon polarization. Clearly the relaxation in the case of Ba$_3$ZnIr$_2$O$_9$ is still strong when a 500Oe field is applied which ensures the dynamical nature of its ground state like a QSOL state.

%\newpage


\begin{thebibliography}{99}
\bibitem{soc_phasedia} Witczak-Krempa, W., Chen, G., Kim, Y. B. \& Balents, L. {\it Annu. Rev. Condens. Matter Phys.} {\bf 5}, 57 (2014).
\bibitem{kim_prl}Kim, B. J. {\it et. al.} {\it Phys. Rev. Lett.} {\bf 101}, 076402 (2008).
\bibitem{noh_prl}Moon, S. J. {\it et. al.} {\it Phys. Rev. Lett.} {\bf 101}, 226402 (2008).
\bibitem{arima_science}Kim, B. J. {\it et. al.} {\it Science} {\bf 323}, 1329 (2009).
\bibitem{Matsuura} Matsuura, H. \& Miyake, K. {\it J. Phys. Soc. Jpn.} {\bf 82}, 073703 (2013).
\bibitem{jce}Rublo, J. \& Perez, J. J. {\it J. Chem. Education} {\bf 63}, 476 (1986).
\bibitem{khaliullin_prl}Khaliullin, G. {\it Phys. Rev. Lett.} {\bf 111}, 197201 (2013).
\bibitem{Cao_prl}Cao, G. {\it et. al.} {\it Phys. Rev. Lett.} {\bf 112}, 056402 (2014).
\bibitem{cava_jssc}Bremholm, M., Dutton, S. E., Stephens, P. W. \& Cava, R. J. {\it J. Solid State Chem.} {\bf 184}, 601 (2011).
\bibitem{FeRu_own} Middey, S. {\it et. al.} {\it Phys. Rev. B} {\bf 83}, 144419 (2011).
\bibitem{fullprof} Rodr\'{i}guez-Carvajal, J. {\it Physica B} {\bf 192}, 55 (1993).
\bibitem{vasp1} Kresse, G. \& Hafner, J. {\it Phys. Rev. B} {\bf 47}, 558 (1993).
\bibitem{vasp2} Kresse, G. \& Furthm\"{u}ller, J. {\it Phys. Rev. B} {\bf 54},11169 (1996).
\bibitem{PAW1} Bl\"{o}chl, P. E. {\it Phys. Rev. B} {\bf 50}, 17953 (1994).
\bibitem{PAW2} Kresse, G. \& Joubert, D. {\it Phys. Rev. B} {\bf 59}, 1758 (1999).
\bibitem{Ir_4f}Otsubo, T., Takase S. \& Shimizu, Y. {\it ECS Transactions} {\bf 3 (1)} 263 (2006).


\bibitem{Matsuura2}Matsuura, H. \& Ogata, M. {\it J. Phys. Soc. Jpn.} {\bf 83}, 093701 (2014).
\bibitem{kt} Hayano, R. S., Uemura, Y. J., Imazato, J., Nishida, N., Yamazaki, T. \& Kubo, R. {\it Phys. Rev. B} {\bf 20}, 850 (1979).
\bibitem{karen} Keren, A. {\it Phys. Rev. B} {\bf 50}, 10039 (1994).
\bibitem{Bouvier}Bouvier, M., Lethuillier, P. \& Schmitt, D. {\it Phys. Rev. B} {\bf 43}, 13137 (1990).
\bibitem{gaplessHC}Han, T. H., Chisnell, R., Bonnoit, C. J., Freedman, D. E., Zapf, V. S., Harrison, N., Nocera, D. G., Takano, Y. \& Lee, Y. S. arXiv:1402.2693
\bibitem{spinon}Clark, L., {\it et. al.} {\it Phys. Rev. Lett.} {\bf 110}, 207208 (2013).
\bibitem{spinon_paramagnet}Yamashita, S., Yamamoto, T., Nakazawa, Y., Tamura, M. \& Kato, R. {\it Nat. Commun.} {\bf 2}, 275 (2011).
\bibitem{spinon3}Yamashita, S., Nakazawa, Y., Oguni M., Oshima, Y., Nojiri, H., Shimizu, Y., Miyagawa, K., \& Kanoda, K. {\it Nat. Phys.} {\bf. 4}, 459 (2008).
\bibitem{gamma_vacancy}Schliesser, J. M. \& Woodfield, B. F. {\it Phys. Rev. B} {\bf 91}, 024109 (2015).
\bibitem{cusb2}Zhou, H. D. {\it et. al.} {\it Phys. Rev. Lett.} {\bf 106}, 147204 (2011).
\bibitem{nisb2}Cheng, J. G. {\it et. al.} {\it Phys. Rev. Lett.} {\bf 107}, 197204 (2011)
\bibitem{niga2}Nakatsuji, S. {\it et. al.} {\it Science} {\bf 309}, 1697 (2005).
\bibitem{mahajan2012}Dey, T. {\it et. al.} {\it Phys. Rev. B} {\bf 86}, 140405(R) (2012).
\bibitem{ramirez}Ramirez, A. P., Hessen, B. \& Winklemann, M. {\it Phys. Rev. Lett.} {\bf 84}, 2957 (2000).






\end{thebibliography}

\begin{thebibliography}{10}
 \bibitem{Matsuura}Matsuura, H., \& Miyake, K. Effect of spin–orbit interaction on (4$d$)$^3$ - and (5$d$)$^3$ - based transition metal oxides. {\it J. Phys. Soc. Jpn.} {\bf 82}, 073703 (2013).

 \bibitem{vasp1}Kresse, G. \& Hafner, J. {\it Ab initio} molecular dynamics for liquid metals. {\it Phys. Rev. B} {\bf 47}, 558 (1993).

 \bibitem{vasp2} G. Kresse \& J. Furthm\"{u}ller, Efficient iterative schemes for {\it ab initio} total-energy calculations using a plane-wave basis set. {\it Phys. Rev. B} {\bf 54}, 11169 (1996).

 \bibitem{HubbardU}Anisimov, V. I., Zaanen, J. \& Andersen, O. K. Band theory and Mott insulators: Hubbard $U$ instead of Stoner $I$. {\it Phys. Rev. B} {\bf 44}, 943 (1991).

 \bibitem{PAW1}Bl\"{o}chl, P. E. Projector augmented-wave method. {\it Phys. Rev. B} {\bf 50}, 17953 (1994).

 \bibitem{PAW2}Kresse, G. \& Joubert, D. From ultrasoft pseudopotentials to the projector augmented-wave method. {\it Phys. Rev. B} {\bf 59}, 1758 (1999).
\end{thebibliography}
\end{document}